\documentclass[11pt]{article}

\usepackage{graphicx}
\usepackage{amssymb}
\usepackage{amsmath,amscd}
\usepackage{url}
\usepackage{citesort}

\def\pa{\partial}
\def\fr{\frac}

\def\ii{\textrm i}
\def\ee{\textrm e}

\newcommand{\db}{de$\,$Broglie}

\newcommand{\dbb}{de$\,$Broglie-Bohm}

\begin{document}

\vspace*{1.0cm}
\noindent
{\bf
{\large
\begin{center}
Pilot-wave approaches to quantum field theory
\end{center}
}
}

\vspace*{.5cm}
\begin{center}
Ward Struyve{\footnote{Postdoctoral Fellow FWO.}}\\
Institute of Theoretical Physics, K.U.Leuven,\\
Celestijnenlaan 200D, B--3001 Leuven, Belgium.{\footnote{Corresponding address.}}\\
Institute of Philosophy, K.U.Leuven,\\
Kardinaal Mercierplein 2, B--3000 Leuven, Belgium.\\
E--mail: Ward.Struyve@fys.kuleuven.be.
\end{center}

\begin{abstract}
\noindent
The purpose of this paper is to present an overview of recent work on pilot-wave approaches to quantum field theory. In such approaches, systems are not only described by their wave function, as in standard quantum theory, but also by some additional variables. In the non-relativistic pilot-wave theory of \db\ and Bohm those variables are particle positions. In the context of quantum field theory, there are two natural choices, namely particle positions and fields. The incorporation of those variables makes it possible to provide an objective description of nature in which rather ambiguous notions such as `measurement' and `observer' play no fundamental role. As such, the theory is free of the conceptual difficulties, such as the measurement problem, that plague standard quantum theory.
\end{abstract}

\section{Introduction}
The pilot-wave theory of \db\ and Bohm for non-relativistic system describes point-particles in physical 3-space, which move under the guidance of the wave function. This theory reproduces the predictions of standard quantum theory (at least in so far those are unambiguous), given a state of equilibrium. Here we are concerned with generalizations of this theory to quantum fields. The purpose is to give an overview of our recent work and to put it into context. Much of this overview is based on \cite{struyve10}, which contains a review and further development of field based approaches, on \cite{struyve06,struyve07b} (with H.\ Westman) which deals with a minimalist approach and on \cite{colin07} (with S.\ Colin) which deals with a Dirac sea approach.

In section \ref{nonrelativistic}, we first recall the pilot-wave theory of \db\ and Bohm for non-relativistic systems and its extension to spin by Bell. Then we turn to quantum field theory. First, in section \ref{fields}, we consider field based approaches. We present Bohm's approach to bosonic fields and discuss the problems with similar approaches to fermionic fields. We also present the minimalist approach, which avoids these problems. Then, in section \ref{particles}, we consider particle based approaches: Bell's lattice approach, a continuum approach in terms of particles and anti-particles by D\"urr {\em et al.}\ and the Dirac sea approach.

\section{Non-relativistic pilot-wave theory}\label{nonrelativistic}
\subsection{Pilot-wave theory of \db\ and Bohm for spinless particles} 
Non-relativistic pilot-wave theory, which was first proposed by \db\ in the 1920s \cite{debroglie28}, and rediscovered by Bohm in the 1950s \cite{bohm52a,bohm52b}, is about point particles in physical 3-space, with positions ${\bf x}_1(t), \dots,{\bf x}_n(t)$, which are guided in their motion by the wave function $\psi(x,t)=\psi({\bf x}_1, \dots,{\bf x}_n,t)$ (for detailed expositions, see \cite{bohm93,holland93b,durr09}). The wave function satisfies the non-relativistic Schr\"odinger equation
\begin{equation}
\ii \hbar \frac{\pa \psi(x,t)}{\pa t}  =  \left( -\sum^n_{k=1} \frac{\hbar^2 }{2m_k}\nabla^2_k + V(x) \right)  \psi(x,t) \,.
\label{1}
\end{equation}
The possible particle trajectories are solutions of the guidance equations 
\begin{equation}
\fr{d {\bf x}_k(t)}{dt} = \frac{\hbar}{m_k} {\textrm{Im}} \frac{{\boldsymbol {\nabla}}_k \psi(x,t)}{\psi(x,t)} \Bigg|_{x=x(t)} = \frac{1}{m_k} {\boldsymbol {\nabla}}_k S(x,t) \bigg|_{x=x(t)} \,,
\label{2}
\end{equation}
where $S$ is the phase of the wave function, that is, $\psi = |\psi|\ee^{\ii S/\hbar}$.

The particle dynamics has the important property that it preserves the density $|\psi|^2$. That is, if a distribution of particles, all guided by the same wave function $\psi$, is given by $|\psi(x,t_0)|^2$ at a certain time $t_0$, then the distribution is given by $|\psi(x,t)|^2$ at any other time $t$. This property is called equivariance. It essentially follows from the continuity equation for $|\psi|^2$,
\begin{equation}
\frac{\pa |\psi|^2}{\pa t} + \sum^n_{k=1}{\boldsymbol {\nabla}}_k \cdot \left( \frac{{\boldsymbol {\nabla}}_k S}{m_k} |\psi|^2 \right) = 0  \,,
\label{2.1}
\end{equation}
which itself follows from the Schr\"odinger equation. 

The distribution $|\psi|^2$ plays the role of an equilibrium distribution and is called the quantum equilibrium distribution. Given the quantum equilibrium distribution and the fact that measurement results are ultimately recorded in positions of things, like instrument needles, computer outprint, etc., it almost follows immediately that the \dbb\ theory reproduces the standard quantum mechanical predictions.

Note that the equivariance of the equilibrium distribution $|\psi|^2$, which plays an important role in showing that the theory reproduces the standard quantum predictions, results from the fact that the velocity field is actually given by $ {\bf j}^\psi_k/|\psi|^2$, where ${\bf j}^\psi_k = \hbar {\textrm{Im}} (\psi^* {\boldsymbol {\nabla}}_k \psi) / m_k$ is the usual quantum current which appears in the continuity equation for $|\psi|^2$. For other quantum theories, in particular quantum field theories, one can develop a pilot-wave theory by introducing a velocity field of the same form (see \cite{struyve09a} for the construction of such velocity fields for arbitrary Hamiltonians).

\subsection{Treatment of spin}\label{spin}
As shown by Bell, the theory can straightforwardly be extended to cover particles with spin \cite{bell66,bell71}. For example, a non-relativistic particle with spin-1/2 is then described by a spinor $\psi_a({\bf x})$ ($a=1,2$), which satisfies the non-relativistic Pauli equation, and a position ${\bf x}$, which evolves according to the guidance equation
\begin{equation}
\fr{d {\bf x}(t)}{dt} = \frac{\hbar}{m} {\textrm{Im}} \frac{ \sum_a \psi^*_a ({\bf x},t){\boldsymbol {\nabla}} \psi_a ({\bf x},t)}{\sum_a |\psi_a({\bf x},t)|^2} \Bigg|_{{\bf x}={\bf x}(t)}  \,.
\label{3}
\end{equation}
The quantum equilibrium distribution is given by $\rho({\bf x},t) = \sum_a |\psi_a({\bf x},t)|^2$.

No extra variables are introduced to represent spin. There exist alternative approaches that do introduce such variables. For example, in one such approach \cite{dewdney86}, the spin-vector ${\bf s}$ is introduced, which is defined as
\begin{equation}
{\bf s}(t) = \frac{\sum_{a,a'}\psi^*_a({\bf x},t) \frac{\hbar }{2}{\boldsymbol \sigma}_{aa'} \psi_{a'}({\bf x},t)}{\sum_a |\psi_a({\bf x},t)|^2} \Bigg|_{{\bf x}={\bf x}(t)} \,,
\label{4}
\end{equation}
where ${\boldsymbol \sigma}$ are the Pauli matrices and where the function on the right hand side is evaluated for the actual particle position ${\bf x}(t)$. As such, at each time, the spin vector is completely determined by the wave function and the actual particle position. 

However, such extra variables, which seem rather artificial, are not necessary. Because results of measurements are generally recorded in positions of things, the theory reproduces the standard quantum predictions (in quantum equilibrium). For example, in a spin measurement with a Stern-Gerlach device, the measured spin of the particle will be up or down depending on whether the particle is detected in the upper or lower half of the detecting screen.

\begin{figure}
\begin{center}
\includegraphics{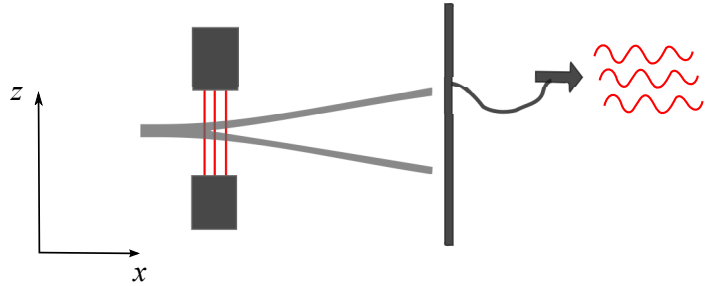}
\end{center}
\caption{\label{sterngerlach}Setup of the Stern-Gerlach experiment. One can identify the Stern-Gerlach magnets, the corresponding magnetic field (represented by the three straight lines), the detecting screen and the pointer that is connected to the screen. The wavy lines represent the electromagnetic radiation surrounding the pointer.}
\end{figure}

\begin{figure}
\begin{center}
\includegraphics{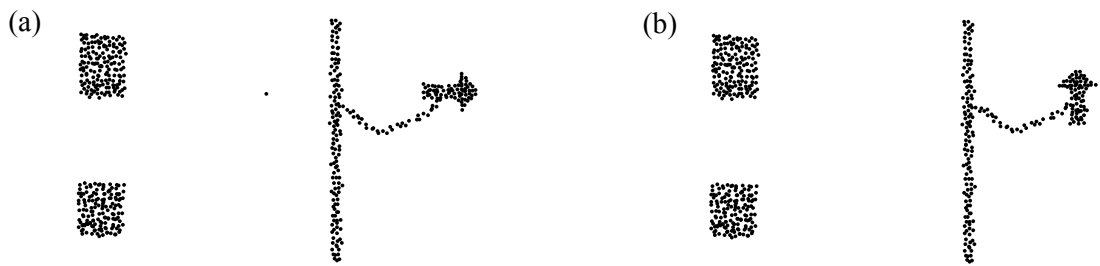}
\end{center}
\caption{\label{dbb}The Stern-Gerlach experiment according to Bell's model for non-relativistic particles with spin. (a) displays the situation just before the particle hits the detecting screen. (b) displays the situation just after the measurement.}
\end{figure}

To conclude this section, we consider the Stern-Gerlach experiment in a bit more detail. Throughout the paper we will give an impression of the various ontologies by sketching the situation for the Stern-Gerlach experiment. The details of the setup are as follows (see Figure \ref{sterngerlach}). We consider a source emitting electrons with spin up in the $x$-direction (meaning that their wave function is an eigenstate of the spin operator in the $x$-direction with eigenvalue $1$). The particles pass through an inhomogeneous magnetic field with decreasing strength in the positive $z$-direction and end up at a detecting screen. The screen is connected to a pointer. In its steady state, the pointer points to the right. After the measurement, it will point up or down, depending on whether the particle was detected in the upper or lower half of the detecting screen. This pointer, which is absent in the usual setup, is introduced merely for convenience. In the usual setup, the incoming particle leaves an imprint on the detecting screen as a result of a chemical reaction, but this is much harder to display in our sketches of the ontology than the orientation of a pointer. For later purposes, we have also drawn the radiation near the pointer. This comprises the thermal radiation emitted by it and the radiation, such as visible light, that is scattered off it. The rest of the environment is ignored. In Figure \ref{dbb}, we sketched the situation of the Stern-Gerlach spin measurement according to Bell's model.

\section{Field ontologies for quantum field theory}\label{fields}
For quantum field theory, there are approaches in terms of particles and in terms of fields. We first consider field approaches. The particle approaches will be discussed in the next section.

The theories that we will consider are not Lorentz invariant at the fundamental level. First of all, there might be regulators such as ultra-violet momentum cut-offs (sometimes implicitly assumed), which violate the Lorentz invariance already at the level of the wave equation. Second, the non-locality as implied by the violation of the Bell inequalities, is hard to combine with Lorentz invariance. In quantum equilibrium, the theories reproduce the predictions of the regularized quantum theory. Hence, insofar the latter are compatible with Lorentz invariance, so will be the equilibrium predictions of the corresponding pilot-wave theories. For an overview of possible ways to achieve Lorentz invariance also at the fundamental level, see \cite{tumulka06}. 

In the following units are assumed for which $\hbar=c=1$.

\subsection{Field configurations for bosonic quantum fields}
Already in his 1952 paper \cite{bohm52b}, Bohm presented a pilot-wave approach to quantum field theory. In this approach, which he considered for the free quantized electromagnetic field, there is an actual field configuration ${\bf A}^T({\bf x})$, corresponding to the transverse part of the vector potential, which satisfies the guidance equation
\begin{equation}
\frac{\partial {\bf A}^T ({\bf x},t)}{\partial t} = \frac{\delta S ({\bf A}^T,t)}{\delta {\bf A}^T({\bf x})}\Bigg|_{{\bf A}^T({\bf x}) = {\bf A}^T ({\bf x},t)} \,,
\label{5}
\end{equation}
where $S$ is phase of the wave functional $\Psi({\bf A}^T,t)$, that is, $\Psi = |\Psi|\ee^{\ii S}$. The wave functional satisfies the functional Schr\"odinger equation
\begin{equation}
\ii\frac{\partial \Psi({\bf A}^T,t)}{\partial t} = \frac{1}{2}  \int d^3 x \left( -\frac{\delta}{ \delta {\bf A}^T({\bf x})} \cdot \frac{\delta}{\delta {\bf A}^T({\bf x})} -  {\bf A}^T({\bf x}) \cdot \nabla^2 {\bf A}^T({\bf x})  \right)\Psi({\bf A}^T,t) \,.
\label{6}
\end{equation}
The quantum equilibrium distribution is given by $|\Psi({\bf A}^T,t)|^2 {\mathcal{D}} {\bf A}^T$. (A suitable regularization, which makes the equations of motion and the quantum equilibrium distribution well defined, is implicitly assumed.)

One can take a similar approach to other bosonic fields (see \cite{struyve10} for a review). But so far this approach has not been successfully applied to fermionic fields. There exist two attempts, one by Holland and one by Valentini, to which we now turn. 

\subsection{Euler angles for fermionic quantum fields}
Holland considered a functional Schr\"odinger representation for the fermionically quantized non-relativistic Schr\"odinger field in which the wave functionals are defined on a space of fields ${\boldsymbol{\alpha}}({\bf k}) = (\alpha({\bf k}),\beta({\bf k}),\gamma({\bf k}))$ in a discretized momentum space \cite{holland881,holland93b}. For each momentum ${\bf k}$, ${\boldsymbol \alpha}({\bf k})$ represents the Euler angles that parametrize the group $SU(2)$. Holland introduced an actual field configuration ${\boldsymbol{\alpha}}({\bf k})$ which is guided by the wave functional $\Psi({\boldsymbol{\alpha}},t)$. However, this configuration lives in momentum space and not in physical space, and Holland did not specify an ontology in physical space. Such an ontology could be constructed from the configuration ${\boldsymbol \alpha}({\bf k})$ and the wave functional (in a way similar to that of the construction of the spin vector for non-relativistic particles in section \ref{spin}), see \cite{struyve10}. However, the empirical adequacy of those theories is very hard to check. 

In \cite{struyve10}, we considered an alternative approach, which employs the sames techniques, but which is easier to analyse. In this approach, the actual configuration is given by a set of Euler angles ${\boldsymbol \alpha}({\bf x}) = (\alpha({\bf x}),\beta({\bf x}),\gamma({\bf x}))$ at each point ${\bf x}$ of a spatial lattice. The configuration is guided by a wave functional $\Psi({\boldsymbol \alpha},t)$. It was found that this model seems inadequate because it does not seem to give an adequate image of general macroscopic matter distributions. In order to see this, we considered the wave functional representing a region filled with ordinary matter and the wave functional representing an empty region. We compared the typical field configurations for those wave functionals. They should typically be distinct in order to be able to distinguish matter from empty space. For a cubic region with edge of length $L$, a lattice spacing $a$ and a matter density $\rho$, we found that they are typically distinct if $L a \rho^{2/3}\gg 1$. For a lattice spacing given by the Planck length, that is $a=10^{-35}$m, and a density $\rho = 10^{30}/$m$^3$ (which corresponds to one particle per atomic distance cubed), $L$ should be much bigger than $10^{15}$m, which is clearly unacceptable. However, since we are dealing with a non-relativistic field theory, a much bigger lattice spacing could be considered. But even for a lattice spacing of the order of $10^{-15}$m, the length $L$ should be much bigger then $10^{-5}$m, which is still too large. There also seems to be very little hope that applying this approach to relativistic quantum field theory will yield a better result. For example, for the quantized Dirac field, the configuration will be given by four sets of Euler angles at each point of the spatial lattice (one for each component of the Dirac spinor), but the lattice spacing should be taken much smaller.

\subsection{Grassmann fields for fermionic quantum fields}
Valentini explored a different approach to fermionic fields \cite{valentini92,valentini96}. His approach starts from a functional Schr\"odinger representation in which the wave functionals are defined on a space of Grassmann fields (that is, anti-commuting fields). The actual configurations are Grassmann fields, which are guided by the wave functional. However, there is a difficulty with this approach. While this functional Schr\"odinger picture bears some resemblance to the functional Schr\"odinger picture for bosonic fields, there are important differences. One is that the wave functionals are not complex valued, but take values in a Grassmann algebra. As a result of these differences, Bohm's approach to bosonic fields does not immediately carry over. The main difficulty is that there does not seem to be a natural measure on the space of Grassmann fields that could serve as the quantum equilibrium measure (also in the standard quantum field theoretical picture, integrals are defined algebraically, instead of in terms of measure theoretical notions). As such, it is not clear what possible guidance equations to adopt, as those are usually found by requiring that they preserve the quantum equilibrium distribution.

\subsection{A minimalist approach: only configurations for the bosonic fields}
In \cite{struyve06,struyve07b}, Westman and the present author considered a radically minimalist approach in which field configurations are introduced only for the bosonic degrees of freedom and no variables for the fermionic ones. This was illustrated for the case of scalar quantum electrodynamics. The wave functional, written as $\Psi_f({\bf A}^T,t)$, is then a functional on the space of fields ${\bf A}^T({\bf x})$, just as in the case of the free electromagnetic field, but it carries a label $f$ representing the fermionic degrees of freedom. (We treat $f$ as discrete, but it could also represent a continuous label.) The time evolution of the wave functional is governed by the appropriate Schr\"odinger equation which involves the Hamiltonian for scalar quantum electrodynamics. 

An actual field configuration ${\bf A}^T({\bf x})$ is introduced, just as in Bohm's approach to the free electromagnetic field, but no variables corresponding to the fermionic degrees of freedom. The guidance equation for this field is
\begin{equation}
\frac{\partial {\bf A}^T ({\bf x},t)}{\partial t} = {\textrm{Im}} \frac{\sum_f \Psi^*_f({\bf A}^T,t) \frac{\delta }{\delta {\bf A}^T({\bf x})} \Psi_f({\bf A}^T,t) }{\sum_f |\Psi_f({\bf A}^T,t)|^2}\Bigg|_{{\bf A}^T({\bf x}) = {\bf A}^T ({\bf x},t)} \,.
\label{7}
\end{equation}
The quantum equilibrium density is given by $\rho({\bf A}^T,t)=\sum_f |\Psi_f({\bf A}^T,t)|^2$.

While there are no variables representing matter, the wave functional which guides the field ${\bf A}^T({\bf x})$ still contains the fermionic degrees of freedom. As such, the field configuration ${\bf A}^T({\bf x})$ will in certain cases behave as if there was matter present. For example, it might look like radiation that has been scattered off some matter distribution, or like thermal radiation emitted by such a distribution. In this way, it was argued that the model is empirically adequate, because there will be an image of macroscopic matter distributions in the radiation field. Nevertheless such a model seems rather far removed from our everyday experience of the world and probably takes minimalism too far.

\begin{figure}
\begin{center}
\includegraphics[width=\textwidth]{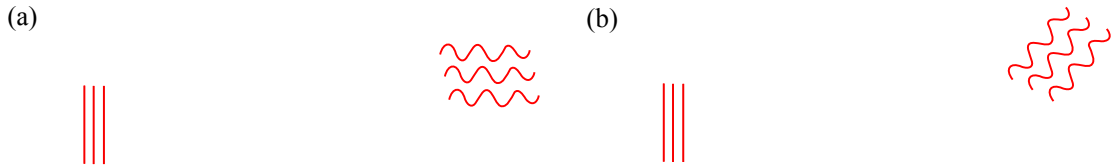}
\end{center}
\caption{\label{minimalist}The Stern-Gerlach experiment according to the minimalist pilot-wave model. There is only the transverse vector potential for the electromagnetic field. The field produced by the magnets and the field surrounding the pointer are displayed. (a) and (b) respectively display the situation just before and just after the measurement.}
\end{figure}

In Figure \ref{minimalist}, the Stern-Gerlach experiment is displayed for this approach. Only the main contributions to the field are drawn. There is the magnetic field induced by the magnets and the radiation field near the pointer. The latter will encode and display the orientation of the pointer.

\subsection{The minimalist approach with additional configurations for the fermionic fields}
Starting from the minimalist model, one can actually introduce extra variables representing the fermionic degrees of freedom \cite{struyve07b}. They could be constructed from the wave functional and the actual field configuration ${\bf A}^T ({\bf x},t)$ in a similar way as the spin vector in the case of non-relativistic particles with spin (cf.\ section \ref{spin}). For example, one could introduce an energy density for the matter field, defined by 
\begin{equation}
E({\bf x},t) =  \frac{\sum_{f,f'} \Psi^*_f({\bf A}^T,t) {\widehat E}_{ff'} ({\bf x}) \Psi_{f'}({\bf A}^T,t)  }{\sum_f |\Psi_f({\bf A}^T,t)|^2}\Bigg|_{{\bf A}^T({\bf x}) = {\bf A}^T ({\bf x},t)} \,,
\label{8}
\end{equation}
where ${\widehat E}_{ff'}({\bf x}) = \langle f | {\widehat E}({\bf x}) |f'\rangle$, with ${\widehat E}({\bf x})$ is the energy density operator for the fermionic fields. The functional on the right hand side is evaluated for the actual field configuration ${\bf A}^T ({\bf x},t)$. So the energy density $E({\bf x},t)$ is defined completely in terms of the wave functional $\Psi_f({\bf A}^T,t)$ and the actual configuration. 

One could introduce other configurations, or extra configurations, such as for example a mass or charge density, in a similar way.
 
\begin{figure}
\begin{center}
\includegraphics[width=\textwidth]{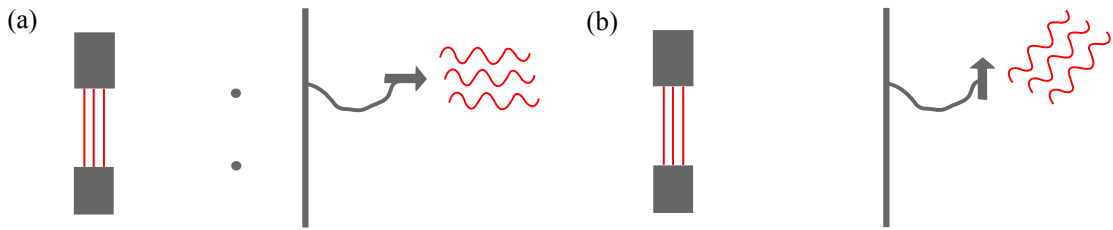}
\end{center}
\caption{\label{minimalistplus}The Stern-Gerlach experiment according to the minimalist pilot-wave model with additional configurations for the fermionic fields. Apart from the vector potential, there is also the energy density of the fermionic fields. (a) and (b) respectively display the situation just before and just after the measurement.}
\end{figure}

In Figure \ref{minimalistplus}, the Stern-Gerlach experiment is displayed for this ontology. We have thereby ignored possible vacuum contributions to the energy density. (Whether there are such contributions probably depends on the exact definition of the energy density operator. This has not been investigated yet.) After the spin-$1/2$ particle has been emitted from the source its energy density is approximately localized. When the density passes the magnets, its splits into two localized packets, one going up and one going down. During the measurement, the energy density will collapse to either packet, depending on the actual value of the radiation field. Similarly, the energy density corresponding to the pointer will display a definite orientation.

\section{Particle ontologies for quantum field theory}\label{particles}
\subsection{Lattice approach in terms of fermion numbers}
Bell was the first to present a pilot-wave approach to quantum field theory in terms of a particle ontology \cite{bell87b}. He considered a spatial lattice and introduced an actual configuration given by the fermion numbers at the lattice sites. No configurations were introduced for the bosonic degrees of freedom. The fermion numbers change in time in a stochastic way, according to a generalized guidance law.  

\subsection{Positions for particles and anti-particles}
D\"urr {\em et al.}\  developed a continuum generalization of Bell's model \cite{durr02,durr031,durr032,durr04} (whereby Bell's lattice regularization is replaced by a suitable regularization compatible with a spatial continuum). They introduced positions for both particles and anti-particles. The trajectories are deterministic, interrupted by stochastic jumps, which usually correspond to creation and annihilation events. This approach has been applied to non-relativistic bosons and fermions, and to Dirac fermions, but not yet to relativistic bosons (the main challenge seems to be to find an appropriate position operator or position POVM which could yield the quantum equilibrium distribution).  

\subsection{Dirac sea approach}
Instead of considering the usual particle--anti-particle picture of standard quantum field theory, as in the approach of D\"urr {\em et al.}, one can also consider the Dirac sea picture. In this picture an anti-particle corresponds to a hole in a sea of negative energy particles. In the corresponding pilot-wave approach, positions are introduced for positive energies, as well as for the negative energies of the Dirac sea. No variables are introduced for the bosons. 

This approach was considered for quantum electrodynamics by Bohm {\em et al.}\ \cite{bohm87b,bohm93} and by Colin \cite{colin03b,colin03c}, who introduced it as a continuum generalization of Bell's model,{\footnote{The reason that Colin obtained a different continuum generalization than D\"urr {\em et al.}\ originates in a different reading of Bell's notion of fermion number. D\"urr {\em et al.}\  understood it as the number of particles plus the number of anti-particles, while Colin understood it as the number of particles minus the number of anti-particles \cite{colin07}.}} and was later extended to the other interactions in the standard model by Colin and the present author \cite{colin07}. 

This approach is deterministic, at least as long as the fermion number is conserved (with fermion number being the number of particles minus the number of anti-particles). The standard model actually predicts a violation of fermion number, but only for very high energies (estimated to be accessible only in accelerators with a performance significantly higher than that of the  Large Hadron Collider \cite{colin07}). The implications for the Dirac sea approach have not yet been studied.

We also considered the empirical adequacy of this approach \cite{colin07}. Since the particles that make up ordinary matter move against a background of particles corresponding to the sea, there might be the worry that the positions do not yield an image of macroscopic matter distributions. Therefore, we calculated how large a region in space should be in order to be able to distinguish a region with ordinary matter from a region without such matter. To simplify the calculation, we ignored the interactions. That is, we assumed that the wave function representing an empty region is given by that of the free vacuum (that is, the state corresponding to no particles and no anti-particles) and that the wave function of the matter is found by applying the particle creation operators on the free vacuum. Since we ignored interactions, we also assumed the physical masses for the particles, instead of the bare masses. A more rigorous analysis, which takes into account the interactions, should deal with the true vacuum and the bare masses of the particles. 

With our simplified analysis, we found that the volume $V$ of the region should satisfy $V \gg (\Lambda / \rho^2)^{3/5}$, where $\Lambda$ is the ultra-violet momentum cut-off and $\rho$ the particle density of the ordinary matter. Taking a cut-off of the order of the Planck scale, that is, $\Lambda \sim 10^{35}/$m, a particle density $\rho \sim 10^{30}/$m$^3$, and a spherical region with radius $b$, we found that the bound is given by $b \gg 10^{-6}$m. This lower bound is rather large. However, it should be noted that the cut-off was taken at the Planck scale. In principle a lower cut-off could be considered. In addition, and more importantly, it is expected that this result will improve by taking into account the interactions. It is expected that the interactions will suppress the fluctuations of the density of background particles (so that they tend to be distributed more uniformly). So far, the only source for suppression of the fluctuations stems from the mutual repulsion of the particles due to the Pauli exclusion principle.

\begin{figure}
\begin{center}
\includegraphics{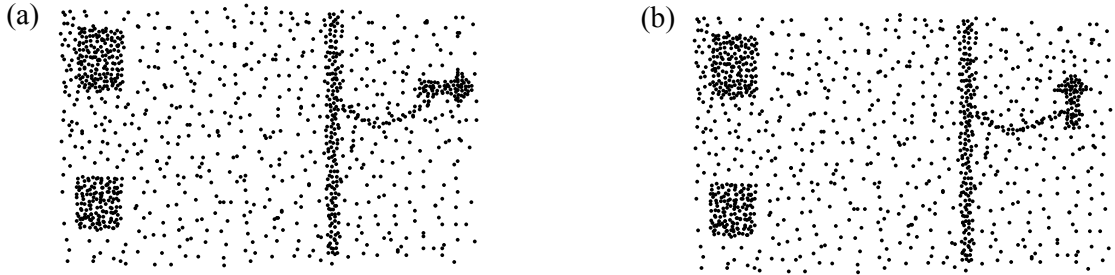}
\end{center}
\caption{\label{diracsea}The Stern-Gerlach experiment according to the Dirac sea pilot-wave approach. (a) and (b) respectively display the situation just before and just after the measurement.}
\end{figure}

In Figure \ref{diracsea}, the Stern-Gerlach experiment is displayed for this approach. Note that the ratio of the particle number density corresponding to a region filled with ordinary matter and the density corresponding to an empty region is not drawn to scale. The ratio is given by $1+8\pi^2 \rho/ \Lambda^3$, which is extremely close to one. Note also that, in these pictures, the electron that was fired to the screen can not be distinguished from the background particles. Maybe it could be distinguished by considering its time evolution. Nevertheless, as we mentioned above, there is good reason to expect that the approach will yield an image of macroscopic matter distributions (like that of the pointer), which is sufficient to account for the standard quantum predictions. 

The approach of D\"urr {\em  et al.}\ actually yields a very similar picture. According to this approach, there are no particles corresponding to the Dirac sea, but instead the particle--anti-particle pairs that are associated to the vacuum fluctuations. While the expected number density of such particles is less than that of Dirac sea particles, it is presumably of the same order (that is, of the order of $\Lambda^3$).

Finally, note that in this approach and in the other approaches considered in this section, one could always introduce field configurations for the bosonic degrees of freedom.

\section{Conclusions}
We described a number of pilot-wave approaches to quantum field theory. Two types of approaches were distinguished: those based on a field ontology and those based on a particle ontology. While the field approach can be applied to bosonic fields, there are problems to extend it to fermionic fields. These problems could be avoided by adopting a radically minimalist approach according to which configurations are introduced only for the bosonic degrees of freedom and nothing for the fermionic degrees of freedom. However, such an approach probably takes minimalism too far. Starting from the minimalist model, there actually is a simple way to introduce field configurations for the fermions, by constructing them out of the bosonic field configurations and the wave function. However, this approach has a rather arbitrary and artificial flavour to it. 

On the other hand, the particle approach seems to work better for fermionic fields. There is the lattice approach by Bell, which is stochastic, and two continuum approaches, namely the particle--anti-particle approach by D\"urr {\em et al.}, which is also stochastic, and the Dirac sea approach, which is deterministic.

\section*{Acknowledgments}
It is a pleasure to thank Samuel Colin for valuable discussions. The author acknowledges support as Postdoctoral Fellow of the FWO-Flanders.

\end{document}